\documentstyle[12pt,a4,epsfig]{article}
\newlength{\abstwidth}
\begin{document}
\newcommand{\am }{{\footnotesize AMPT }}
\newcommand{\amm}{{\footnotesize AMPT}}
\newcommand{\ammm}{{\footnotesize AMPT. }}
\newcommand{\src }{{short-range correlations }}
\newcommand{\srcc }{{short-range correlations }}
\newcommand{\lrc }{{long-range correlations }}
\newcommand{\lrcc }{{long-range correlations}}
\newcommand{\flu }{fluctuations }
\newcommand{\fll}{fluctuations}
\newcommand{\fl}{fluctuations.}
\newcommand{\cor}{correlations }
\newcommand{\corr}{correlations}
\newcommand{\deta}{ $\Delta\eta$ }
\newcommand{\et}{ $\eta$ }
\newcommand{\ett}{ $\eta$}
\newcommand{\dik}{d$_{ik}$ }
\newcommand{\fq}{F$_q$ }
\newcommand{\ft}{F$_2$ }

\thispagestyle{empty}
\begin{center}
\noindent{\Large {\bf A study of event-by-event fluctuations in relativistic heavy-ion collisions}}\\[5mm]
\end{center}

\noindent{Shakeel Ahmad\footnote{email: Shakeel.Ahmad@cern.ch}, M.M. Khan, Shaista Khan, A. Khatun and M. Irfan}\\
{\it Department of Physics, Aligarh Muslim University, Aligarh 202002, India}\\


\begin{center}
{\bf Abstract}\\[2ex]
\begin{minipage}{\abstwidth} A method for selecting events with densely populated narrow regions or spikes in a given data sample is discussed. Applying this method to 200 A GeV/c $^{32}$S-AgBr and $^{32}$S-Gold collision data, a few events having 'hot regions' are chosen for further analysis. The finding reveals that a systematic study of particle density fluctuations, if carried out in terms of scaled factorial moments, and the results are compared with those for the analysis of correlation free Monte Carlo events, would be useful in identifying events with large dynamical fluctuations. Formation of clusters or jet-like structure in multihadronic final states in the selected spiky events is also looked into and compared with the predictions of AMPT and independent emission hypothesis models by carrying out Monte Carlo simulation. The findings suggest that clustering or jet-like algorithm adopted in the present study may also serve as an important tool for triggering different classes of events.\\

\noindent {\it Keywords:} Relativistic heavy-ion collisions, Event-by-event fluctuations, Clusterization, Intermittency.\\
PACS Number(s): 25.75.--q, 25.75.Gz 
\end{minipage}
\end{center}
\newpage
\section{Introduction}
\noindent Any physical quantity measured in an experiment is subject to \fll. These \flu depend on the property of the system and are expected to provide some useful clue about nature of the system under study\cite{1}. As regards relativistic ion-ion (AA) collisions, the system so created is a dense and hot fireball consisting of hadronic and (or) partonic matter\cite{1}. One of the main aims of such a study is to investigate the existence of partonic matter in the early life of the fireball. Study of \flu in relativistic AA collisions helps check the idea that \flu of a thermal system are directly related to the various susceptibilities\cite{1,2} and could usefully serve as an indicator of possible phase transitions. Furthermore, large event-by-event (ebe) \fll, if observed, might be a signal for the presence of a distinct class of events produced  via formation of QGP\cite{2,3,4,5}, as under extreme conditions of energy density and temperature, a novel phase of matter, 'the QGP' is likely to be produced. Therefore, search for occurrence of phase transition from hadronic matter to QGP still remains a favorite topic of interest of high energy physicists\cite{6,7,8}. Collisions between heavier nuclei at relativistic energies are believed to the best site to search for such a phase transition. However, even if such conditions are achieved, not all the events will produced QGP, because it is not yet known whether cross section for QGP formation would be large. Hence, to search for QGP formation, one has to carry out analysis on ebe basis\cite{9}.\\

\noindent A major contribution to the observed \flu results due to finite number of particles used to define an observable in a given event and are referred to as statistical \fl  These \flu can be evaluated by considering the independent emission of particles or by event mixing technique\cite{10,11}. The other \flu present will be of dynamical origin and may be classified into two categories\cite{1}: a) \flu which do no change on ebe basis, for instance, two particle correlations due to Bose-Einstein statistics or due to decays of resonances and b) \flu which vary on ebe basis. Relevant example is charged to neutral particle multiplicity ratio due to creation of regions of DCC(disoriented chiral condensate) or creation of jets which contribute to the high transverse momentum (p$_t$) tail of p$_t$ distributions; DCC is a region in space in which chiral order parameter points in a direction in isospin space, which is different from that favoured by the true vacuum\cite{12}. DCC formation can produce a spectacular event structure within a region of detector dominated by charged pions and the other by neutral pions. This behaviour may have been observed in Centauro events\cite{12,13,14,15}. It will be interesting to know whether there exist some mechanisms through which DCC formation in relativistic AA collisions can be explained. If such mechanisms invoke the occurence of QCD phase transition in an essential way then signals of DCC formation may help conclude QGP formation\cite{12}. Several workers\cite{16,17,18,19,20,21,22} have investigated ebe \flu and concluded that \flu are of dynamical origin. However, most of these investigations are based on data collected from the detectors having limited acceptance. This not only reduces the number of observed secondary particles but may also distort some important event characteristics. Data on collisions of beams of 200A GeV/c $^{32}$S  with AgBr and gold nuclei in emulsion respectively are, therefore analyzed. Needless to emphasize the conventional nuclear emulsion technique has two main advantages over other detectors: (i) its 4\(\pi\) solid angle coverage and (ii) data are free from biases due to full phase space coverage\cite{9,23}. Although the projectile energy in the present experiment is not too high, yet it is expected that the results obtained on various aspects, of ebe \flu and their comparison with the predictions of Monte Carlo model, AMPT, as well as with those obtained from the analysis of correlation free (mixed) events would lead to some interesting results. It should be mentioned that there are indications of phase transition from normal hadronic matter to QGP in $^{207}$Pb -$^{207}$Pb collisions at $\sim$ 30A GeV energy\cite{24,25}.

\section{Experimental Details}
\noindent Two samples of events, produced in the interactions of 200A GeV/c $^{32}$S beam with AgBr and Gold nuclei in emulsion at 200A GeV/c are used in the present study; the numbers of events in the two samples are respectively 452 and 542. These events are taken from the emulsion experiments performed by EMU01 collaboration\cite{26,27,28,29}. In one of the experiments, 12 BR-2 type emulsion stacks consisting of 30 pellicles were horisontally exposed to 200A GeV/c $^{32}$S beam at CERN SPS. The events were searched for by along-the-track scanning procedure, which gives a reliable minimum bias sample because of its inherent high detection efficiency\cite{11,26,27}. The collisions lying within 2-5 cm from the edge of the pellicles were considered for various measurements and subsequent analysis\cite{26,27}. The tracks of the secondary particles were identified\cite{11,26,27,30} on the basis of their ionization. The tracks having ionization, \(I < 1.4I_0\), where \(I_0\) is the minimum ionization produced by a singly charged relativistic particle, are known as shower tracks. The number of such tracks in an event is denoted by n$_s$. The tracks with ionization in the range: \(1.4I_0 \leq I \leq 10I_0\) are termed as grey tracks, while those having ionization \(I > 10I_0\) are referred to as black tracks. The numbers of grey and black tracks produced in an events are denoted by n$_g$ and n$_b$ respectively. The grey and the black tracks are jointly referred to as heavily ionizing tracks and the number of such tracks in an event is represented by n$_h$(= n$_b$ + n$_g$). Events with n$_h$ \(\geq\) 8 are envisaged to be produced exclusively due to interactions with AgBr group of nuclei, whereas those with n$_h$ \(\leq\) 7 are either due to the interactions with H or CNO group of targets or due to the peripheral collisions with AgBr group of nuclei\cite{11,27,30}. On the basis of these criteria, events produced in the interactions of $^{32}$S ions with AgBr of targets were considered for the present analysis. In the other experiment, emulsion chambers were exposed\cite{27,28} to $^{32}$S beam from CERN SPS. For the $^{32}$S-Gold exposures, chambers were additionally equipped upstream, with a gold foil of 250 $\mu$m thickness immediately after two sheets of polysthyrene of 780 $\mu$m thickness each coated with 220 $\mu$m thick emulsion layers on both sides. Various details, like chamber design, methods of measurements and scanning efficiency etc., may be found elsewhere\cite{27,28,31}. In order to compare the results  of the present work with those predicted by  A Multi Phase Transport (AMPT) model, a matching numbers of events equal to the experimental ones are simulated using the code, ampt-v1.21-v2.21\cite{32}. The events are simulated by taking into account the percentage of interactions which occur in collisions of projectile with various targets in emulsion\cite{29}. While generating the \am events, the value of impact parameter for each data sample was so set that the mean multiplicities of the relativistic charged particles, \(<n_s>\) became nearly equal to those obtained for the real data sets, The values of mean multiplicities of relativistic charged particles for the experimental and \am samples are listed in Table~1. The errors in the values are statistical ones and are estimated using standard procedure; the error is calculated using \(\sigma/\sqrt{N_{evt}}\),  where \(\sigma(= \sqrt{<n_s^2> - <n_s>^2})\) is the dispersion of the multiplicity distribution and \(N_{evt}\) denotes the number of events in a given sample. Furthermore, in order to search for the evidence of \flu of dynamical origin, if any, the results are compared with those obtained from the analysis about the data free from the dynamical correlations. The technique of event mixing gives such a reference data sample in which dynamical correlations amongst the particles are completely destroyed. The mixed event samples corresponding to the real and \am data sets are simulated by adopting the standard procedure\cite{11,33}, according to which a mixed event with 'n' number of particles is generated by randomly picking up one particle from each of the 'n' events selected randomly from the original sample. Thus, in a mixed event, there will be no two particles coming from the same event.

\section{Results and Discussion}
\subsection{Presence of high density phase region}
\noindent As already mentioned, even if suitable condition for QGP formation is reached, not all the events would produced QGP search of such rare events from a large sample of events is, therefore, not an easy task. For this purpose, one has to find possible ways to characterize each event which, in turn, may lead to triggering of different classes of events and may help identify anomalous feature. A search is, therefore, carried out for the events with high density phase region, where a lot of entropy is confined within a small domain. These high density regions in one dimensional distributions are usually referred to as 'hot regions' or spikes\cite{9}. For searching spikes or 'hot regions', a parameter, d$_{ik}$ is introduced\cite{9}, where d$_{ik}$ measures local deviation from the average particle density in units of statistical errors. For a given distribution, d$_{ik}$ for i${th}$ bin of k${th}$ event is expressed as:\cite{9}
\begin{eqnarray}
   d_{ik} = \left(n_{ik} - \frac{N_k}{<N>} <n_{ik}>\right)/\sigma_{ik}
\end{eqnarray}

\noindent where \(n_{ik}\) is the charged particle multiplicity in \(i^{th}\)-bin of \(k^{th}\)-event, \(N_k\) is the multiplicity of the \(k^{th}\)-event, \(\sigma_{ik} (= \sqrt n_{ik}\)) denotes the statistical error and \(<N>\) is the mean multiplicity of the sample.\\

\noindent d$_{ik}$ distributions in the pseudorapidity,\et, space for $^{32}$S-AgBr and $^{32}$S-Gold collisions at 200A GeV/c are compared with a reference distributions in Fig.1; reference distribution for each of the two sets of experimental data are obtained by carrying out parallel analyses of the corresponding mixed event samples. The value of \et-bin width is fixed to be 0.2. It may be noted that the \dik distributions for the real data have relatively longer tails in the region of high \dik values. Such a tail is rather more pronounced in the case of $^{32}$S-Gold data. This fact is more clearly reflected in Fig.2, in which the distributions of the differences of the data and mixed events are displayed. This indicates that the experimental data might have some events with 'hot regions' or spikes in the \et space; spikes or 'hot regions' are defined as the regions having relatively larger \dik : The value of \dik, showing a spike has been taken as \(d_{ik} \ge 2.5\) by Cherry et al\cite{9}. In the present study, we have taken the same value of \dik to identify a spike in the case of $^{32}$S-Gold collisions. However, for $^{32}$S-AgBr interactions, the value of d$_{ik}$ is taken to be 2.2. The reason for taking a somewhat smaller value of \dik  is to have a few percent of the total events so that further analysis of this class of events may be statistically reliable. If \dik is taken equal to 2.5 for $^{32}$S-AgBr collisions, only a few events are available for further analysis; this number becomes so small that conclusions drawn shall not be reliable. It has, however, been ensured before considering \dik = 2.2 for $^{32}$S-AgBr interactions that the values of various parameters linked with intermittency and clusterization, to be discussed in the coming section, are found to be almost identical for \dik = 2.2 and 2.5. This fact is clearly reflected in Fig.3, in which ln\ft values are plotted against lnM for  $^{32}$S-AgBr and  $^{32}$S-Gold collisions taking \dik = 2.2 and 2.5. It is noticed from the figure that the variations of ln\ft with lnM for the two \dik cuts are essentially similar. The probability of occurrence of spikes, P(\dik) and the average sizes of the spikes, \(<d_{ik}>\) with \dik $\ge$ 2.5 for $^{32}$S-Gold and \dik $\ge$ 2.2 for $^{32}$S-AgBr collisions are presented in Table~2 for various \et bin widths. The errors are the statistical ones and are estimated as described in section-2. The distributions of \dik for \am and corresponding mixed events are also plotted in Fig.1, while the distributions of differences of \am and mixed events are shown in Fig.2. It may be noted from the figure that the \dik distributions for the \am and the mixed events are almost of identical shapes, particularly in the regions of larger \dik, i.e., \dik $ \ge$ 2.0. The values of P(\dik) and \(<d_{ik}>\) for these data sets are also given in Table~2. Following important inferences may be made from Figs.1 and 2 and Table~2:
\begin{itemize}
\item Occurrence of spikes are rare but their presence can not be ignored, especially in the case of experimental data.
\item For the real data, average sizes of spikes are observed to be larger than those obtained for the mixed event samples. The difference in the \dik values for the real and mixed events are rather more pronounced for larger \et-bin widths.
\item For the \am sample, \dik values for the real data and the corresponding mixed events are nearly the same.
\end{itemize}

\noindent Thus, by studying \dik distributions, the rare events having spikes may be separated from those exhibiting no 'hot regions' for further analysis.\\

\subsection{Factorial Moments}
\noindent The first investigation dealing with intermittent behaviour in multiparticle production at relativistic energies was based on the single JACEE event analysis\cite{34,35} in which unexpectedly large local multiplicity \flu were observed\cite{36}. However, it was soon realized that intermittency analysis can be done using events of any multiplicity provided a proper averaging procedure is adopted\cite{37}. A power law growth of scaled factorial moments (SFMs), \fq,  with decreasing phase space bin width, referred to as intermittency, emerged as a new tool to study the non-linear phenomena in hadronic and nuclear collisions at high energies\cite{34,35,38,39,40}. This method of SFMs has been extensively used\cite{41,42,43,44,45,46,47} to search for non-linear phenomena in high energy hadron-hadron, hadron-nucleus and nucleus-nucleus collisions in a wide range of incident energy. Recently, SFMs have been used to study various processes at SPS and RHIC energies\cite{48} with the aim to scan the phase diagram in a systematic search for the QCD critical point. $^{12}$C-$^{12}$C, $^{28}$Si-$^{28}$Si and $^{207}$Pb-$^{207}$Pb collisions at 158A GeV/c have been studied\cite{49} to search for intermittent \flu in transverse dimensions. The investigations have also been carried out for \(\pi^+\pi^-\) pairs having invariant mass very close to two pion threshold. It has been reported\cite{49} that the power-law \flu in the freeze out state of $^{28}$Si-$^{28}$Si collisions approaches in size as predicted by critical QCD, while for larger systems, like $^{207}$Pb-$^{207}$Pb, this method can not be applied without entering into the invariant mass region with strong coulomb correlations because in such large systems, high multiplicity of the produced pions combined with the restrictions imposed by the necessacity to exclude the coulomb correlations and the resolution of the experiment decrease the sensitivity to the sigma fluctuations near the two pion threshold\cite{49}. The value of intermittency index, $\phi_2$, for such case is found to be vanishingly small. Such a small value of $\phi_2$, whether due to the effect of high multiplicity or to a genuine noncritical nature of the freeze-out state of the system needs to be looked into. This can not be resolved without penetrating the coulomb region to cross 2m$_\pi$ threshold.\\

\noindent It has also been pointed out\cite{34,50,51,52,53} that although there are clear advantages of the averaging procedure adopted in the SFMs studies, yet it may not fully account for all the \flu a system may exhibit and there are chances that some interesting processes may be suppressed which might be present in  a part of events, e.g., the unique properties due to the presence of QGP would manifest only in few events\cite{54}. We have, therefore, analyzed a few spiky events separately to study intermittent \flu and compared  with those obtained from the non-spiky event analysis. The distinct difference between the results from the analyses of spiky and non-spiky events is expected to lead to test for each of the spiky events individually to single out the events of interest.\\

\noindent  \noindent The event factorial moment of order q is defined as\cite{26}
\begin{eqnarray}
F_q^{(e)} = \frac{<n(n-1)....(n-q+1)>_e}{<n>_e^q}
\end{eqnarray}
\noindent where 'n' represents particle multiplicity in a particular pseudorapidity bin, q is the order of moments, while the quantities within angular brackets with subscript e denote the event averaged values.\\

\noindent In order to calculate the values of \ft, all the relativistic charged particle having their \et values in the range \deta(= \(y_c \pm\) 3.0) are considered. This region of \deta is then divided in M cells, each of width \(\delta \eta\) such that the number of cells are equal to \(\Delta \eta/\delta\eta\). The values \(F_q^{(e)}\) for q = 2, \ft , are calculated for spiky and non-spiky events for the experimental, \am and corresponding mixed events. Spiky and non-spiky experimental \am events are sorted out by applying \dik cut using Eq.2; events having spikes with \dik = 2.5 for $^{32}$S-Gold and 2.2 for $^{32}$S-AgBr collisions were taken as spiky events, while those having no spikes with these \dik values were grouped as non-spiky ones. Values of the mean multiplicities of the spiky and non-spiky events for the real and \am events are presented in Table 1. It may be noted from the table that the values of \(<n_s>\) for the non-spiky events are closer to those obtained for all the events of the sample. However, for the spiky events,  \(<n_s>\) is found to be higher in comparison to the one obtained for all the events. This reveals that  spikes occur more often in the events having relatively higher multiplicities. The same procedure was applied for sorting out the spiky and non-spike mixed events, i.e., first the event mixing was done and then the \dik cuts were applied to get the spiky and non-spiky samples. Variations of \ft with the number of cells, M(=2-20), for the experimental and mixed events are displayed in Fig.4. It is interesting to note in Fig.4 that \ft values for the spiky events are significantly larger than those for the non-spiky events, whereas this difference disappears after mixing the events, i.e., the values of \ft for the two categories of events become almost equal. Furthermore, the values of \ft, for the \am events, plotted against the cell size, M, in Fig.5, are seen to be larger for the spiky events in comparison to those for the non-spiky ones. However, the difference between the \ft values for the two types of events is somewhat smaller than those observed for the real data.\\

\noindent As mentioned in the preceding sections, intermittent pattern of the multiplicity \flu results in the power law behaviour of the moments\cite{35,55} of the form:

\begin{eqnarray}
<F_q^{(e)}> = M^{\phi_q} , \hspace{3mm} 0 < \phi_q < q-1
\end{eqnarray}

\noindent where \(\phi_q\) characterizes strength of the intermittency signal. Therefore, a linear dependence of ln\(F_q\) on lnM upto the limit of experimental resolution or the statistical limit is envisaged\cite{55}. The physical significance of \(F_q\) is explained\cite{35} on the basis of self similar cascade model. The values of \(\phi_2\) for various data sets are computed by performing linear fits to the data of the form:

\begin{eqnarray}
ln<F_2> = A- \phi_2lnM
\end{eqnarray}

\noindent where, A is a constant; the fits are performed in the linear regions of the plots, i.e., leaving the first and the last data points. The values of \(\phi_2\) for various data samples are presented in Table 3. The errors associated with $\phi_2$ are statistical ones, estimated from the fitting prodedure. It may be noted from the table that values of \(\phi_2\) are larger for the spiky events in comparison to those for the non-spiky events. Moreover, the values of \(\phi_2\) for $^{32}$S-AgBr and $^{32}$S-Gold collisions are almost the same for the spiky events. In the case of \am data, values of \(\phi_2\) for the spiky events are somewhat smaller in comparison to those for the real data, indicating that the experimental data exhibit larger intermittent \flu than those predicted by the \am model. As far the non-spiky events are concerned, the values of \(\phi_2\) for the various data sets are essentially the same and match with those for the mixed events. It may, therefore be remarked that the method of SFMs seems to be quite suitable for preliminary identification of a distinct class of events showing up significant \flu and thereafter more advanced triggering, like particle ratio, enhanced particle multiplicities in certain kinematical regions, etc, may be applied to the selected data sets. It may be noted that the method of SFMs may be applied to the data at RHIC and LHC energies successfully because at these energies multiplicities of the particles produced in the central collisions are quite high, which may allow to apply suitable p$_t$ cut to draw more meaningful conclusions. Also, due to high multiplicities, method of SFMs can be applied to individual or small sample of rare events, sorted out after applying suitable \dik cuts. \\

\subsection{Clusterization}
\noindent Present analysis reveals, that there are a few events in the experimental data samples which have high particle density regions and still fewer events in the \am  generated samples. Such regions of high particle density are envisaged to arise due to the decays of a heavier clusters or several clusters or jets of relatively smaller sizes\cite{9,56}. In order to confirm whether the observed spikiness are due to some dynamical reasons, we have examined the presence of clusters or jet-like phenomena following the algorithm applied to \(p\overline{p}\) collisions\cite{57}. This algorithm is somewhat different to that adopted in refs.58 and 59 in which formation of clusters and their sizes were looked into by histogramming the pseudorapidity differences between the nth nearest neighbors. The present approach is rather suitable for searching the high density region in \(\eta-\phi\) space, which provides a clean separation in the \(\eta-\phi\) metric in the low multiplicity and low particle density final state\cite{9}. In order to test how the jet algorithm works for high particle density data and to what degree of clustering in two dimensional space, analysis of the spiky and non-spiky events must be carried out separately. The method of clustering is envisaged to help estimate the cluster frequencies and the cluster multiplicities on ebe basis. Since the observables are very sensitive to the total event multiplicities, a comparison between the real and mixed events (with matching multiplicities) findings would yield to some interesting results. A detailed description of the method of analysis, which involves grouping of particles into clusters are given in detail in refs.9,60 and 61. However, considering useful a brief description of the analysis is, therefore, presented below.\\

\noindent For a particle i of an event having 'n' particles, its \(r_{ik}\) value with respect to the next particle k (k $\ne$ i) is calculated using the relation \(r_{ik}=\sqrt(\delta\eta^2 + \delta\phi^2))\), where \(\delta\eta\) and \(\delta\phi\) respectively denote the differences between the pseudo rapidities and azimuthal angles of \(i^{th}\) and \(k^{th}\) particles; this gives a cone of radius \(r_{ik}\) containing \(i^{th}\) and \(k^{th}\) particles. Thus, starting from \(1^{st}\) particle, i.e., i=1, its \(r_{ik}\) value is calculated with respect to \((i+1)^{th}\) particle. If this value is less than a pre-fixed value r, the pair is treated as a cluster of two particles. Once a cluster is obtained another particle is added to it and checked whether it may also be grouped in the same cluster by calculating \(r_{i,i+2}\) value and comparing with the pre-fixed value r. A cluster is treated to be genuine if it has at least 'm' particles with m$\ge$2. Once a cluster is obtained, another cluster is searched for using the remaining particles of the event. It is however, ensured that once a particle is assigned to a particular cluster, it is not again considered for inclusion in the next cluster. Using this approach, the following parameters are calculated for a given value of r:
\begin{itemize}
\item Number of clusters in each event with each clusters having at least 'm' particles.
\item Number of particles in a cluster.
\end{itemize}
\noindent It should be noted that for a very small value of r,  there may be no or only a few clusters in an event, whereas, for a very large value of r, almost all the particles of an event will fall into a single large cluster. The number of clusters in an event with (m$\ge$5) and average number of particles in a cluster are calculated for various data sets. Variations of mean cluster multiplicity $<m>$ and mean number of clusters,\(<n_{cl}>\) in an event with \(r^2\) are plotted in Figs.6 and 7 for the experimental and mixed events. Similar plots for the AMPT events are displayed in Figs.8 and 9. The following observations may be made from Figs.6-9.\\
\begin{enumerate}
\item For Mixed Events: Figs.8 and 9, mean cluster multiplicity, $<m>$ increases from $\sim$  6 to 10 for $^{32}$S-AgBr and to 20 for $^{32}$S-Gold collisions. The data points for spiky and non-spiky events overlap and have the same patterns of variations. Variations of \(<n_{cl}>\) with r show almost the same ("quiet") pattern for the two categories, spiky and non-spiky events having broader maxima for \(r^2\) between 0.2-0.3 and thereafter decreases slowly with increasing \(r^2\) values. The maximum values of mean numbers of clusters are found to be 20 and 40 respectively for the collisions due to AgBr and Gold targets. \\

\item For Real Events: Values of $<m>$ for the spiky events are much higher than those for the non-spiky events, indicating that clustering effects dominates in the case of spiky events from \(<n_{cl}>\) vs \(r^2\) plots it is noticed that relatively more number of clusters are in the case of spiky events as compared to those for non-spiky events. It is interesting to note that the dominant clustering effect in the spiky events is a distinct feature of the data which disappears after events are mixed.\\

\item For AMPT Events: The trends of variations of $<m>$ and \(<n_{cl}>\) with \(r^2\) are essentially similar to that observed for the real data but with smaller magnitude. The finding thus suggests that AMPT model also predicts particle production through formation of clusters; more clusters of larger sizes being formed in the case of spiky events.\\

\end{enumerate}                          

\noindent A comparison of the results for the experimental and AMPT event samples reveals that even in the case of non-spiky events more clusters of rather larger sizes are produced in the case of real data in comparison to those observed for the AMPT events. These findings, thus, indicate that the algorithm used in the present study for investigating clusters or jet-like phenomena seems to be quite suitable to sort out the events having 'hot-regions' or spikes. A comparison of the results for the real and AMPT data with those for the mixed event samples indicates that there is a dominant clustering effect present in the real and AMPT data and these event samples do not agree with the hypothesis of independent particle emission.

\section{Conclusions}
\noindent Analysis of $^{32}$S-AgBr and $^{32}$S-Gold collisions at 200A GeV/c is carried out on an ebe basis and the results are compared with those for the mixed and AMPT event samples. Enhanced particle densities in the narrow {$\eta$} bins are searched for by studying the \(d_{ik}\) distributions in different {$\eta$}-bin widths and compared with the reference distributions due to the mixed events. The findings reveal that there are a few events present in the real data which have spiky or high particle density regions in the {$\eta$} space. Such regions of high density are also observed in the case of AMPT events but with somewhat smaller magnitudes. These spiky events sorted from the real and AMPT data are separately analysed to explore some suitable method for triggering different  classes of events.\\
\noindent The method of scaled factorial moments, when applied to spiky and non-spiky events, indicates that there is a marked difference in the values of \(F_2\) obtained for the two classes of events.\\A dominant cluster or jet-like phenomenon in two  dimensional {$\eta$}-{$\phi$} space is observed in the case of spiky events. Dominance of clusterization is noticed to disappear in the case of mixed events. The findings also reveal that clustering effects are significantly stronger in the case of experimental data; AMPT data also exhibit clustering effect which is somewhat less dominant in comparison to that observed for the real data. The findings, thus, suggest that the real as well as AMPT data do not agree with the hypothesis of complete independent emission. Although analysis involving intermittenccy and clusterization, carried out in the present study, may not lead to some definite conclusions due to limited statistics, yet it may be remarked that the method adopted in the present study may be used as a tool to  select special class of events having high particle density regions, for further analysis to search for the signals of formation of some exotic states like DCC or QGP. So for as the collision data at RHIC and LHC energies are concerned, it may be noted that by applying suitable \dik cuts, rare spiky events can be sorted out for further studies; analysis of these individual rare events would be statistically reliable for number of particles in  such events will be significantly high.

\section{Acknowledgments}
\noindent  The authors are thankful to Prof. A. Bhasin, Jammu University, Jammu for providing the raw data and also for several fruitful discussions S.Ahmad had with her.

\newpage
\begin{figure}[th]
\centerline{\psfig{file=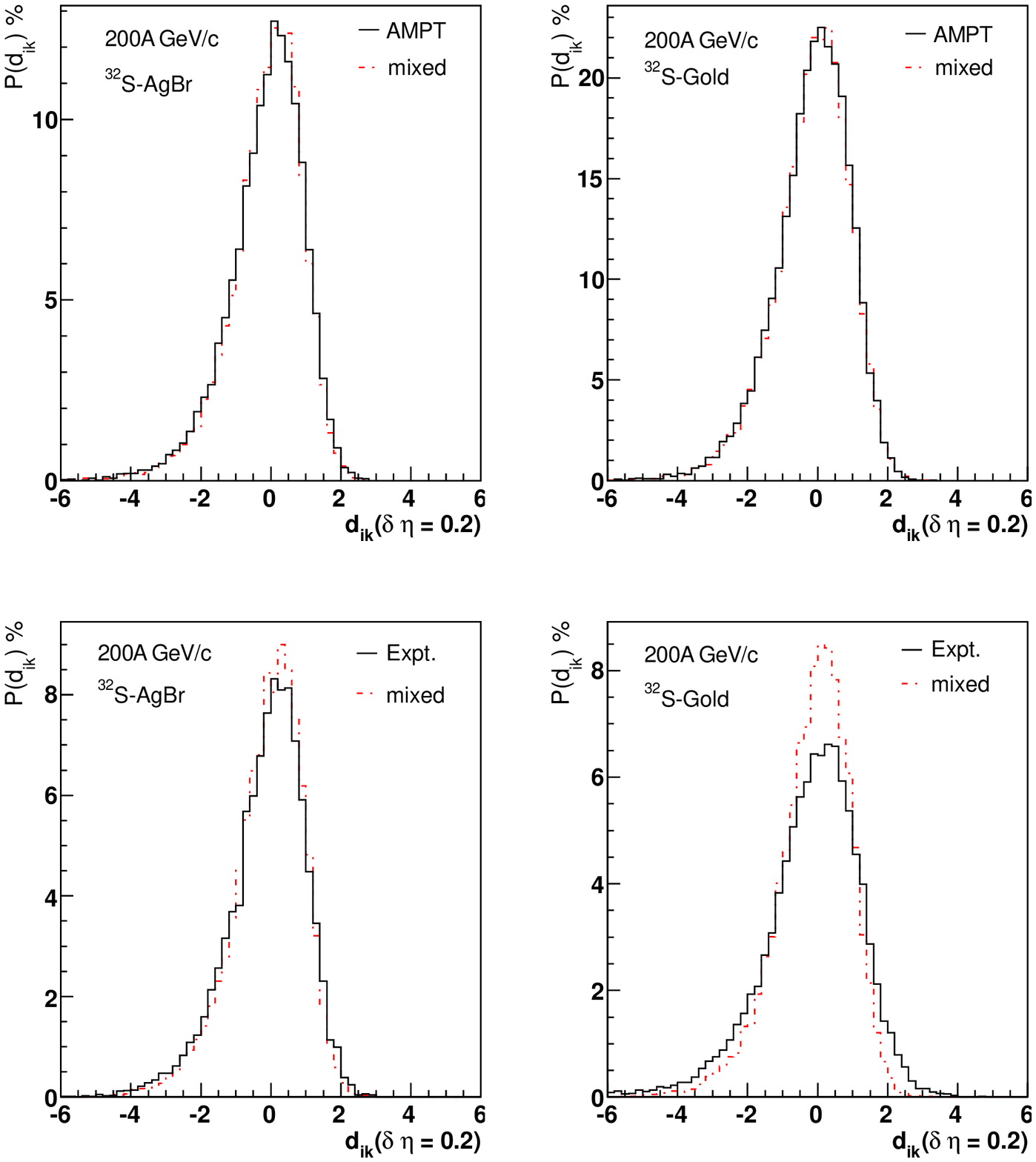,width=8cm}}
\vspace*{8pt}
\caption{\dik distributions for the experimental and \am events compared with the mixed events.}
\end{figure}

\newpage
\begin{figure}[th]
\centerline{\psfig{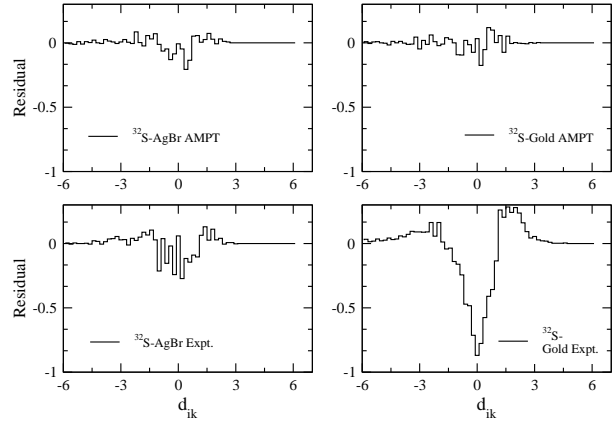}}
\vspace*{8pt}
\caption{The residual distributions between the real/\am data and corresponding mixed event \dik spectra.}
\end{figure}

\newpage
\begin{figure}[th]
\centerline{\psfig{file=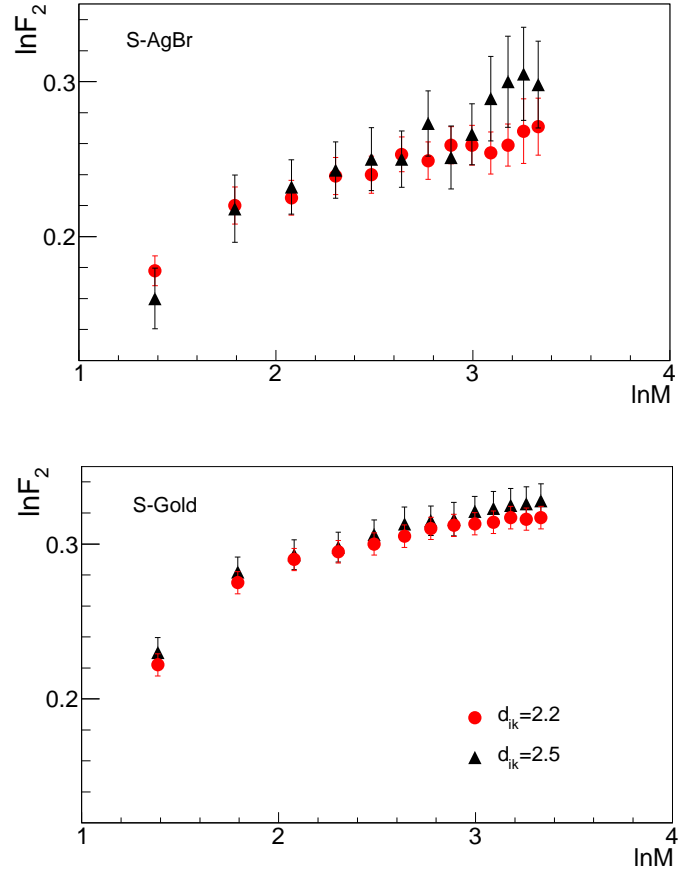,width=10cm}}
\vspace*{8pt}
\caption{Variations of lnF$_2$ with lnM for $^{32}$S-AgBr and $^{32}$S-Gold collisions for \dik cuts = 2.2 and 2.5.}
\end{figure}

\newpage
\begin{figure}[th]
\centerline{\psfig{file=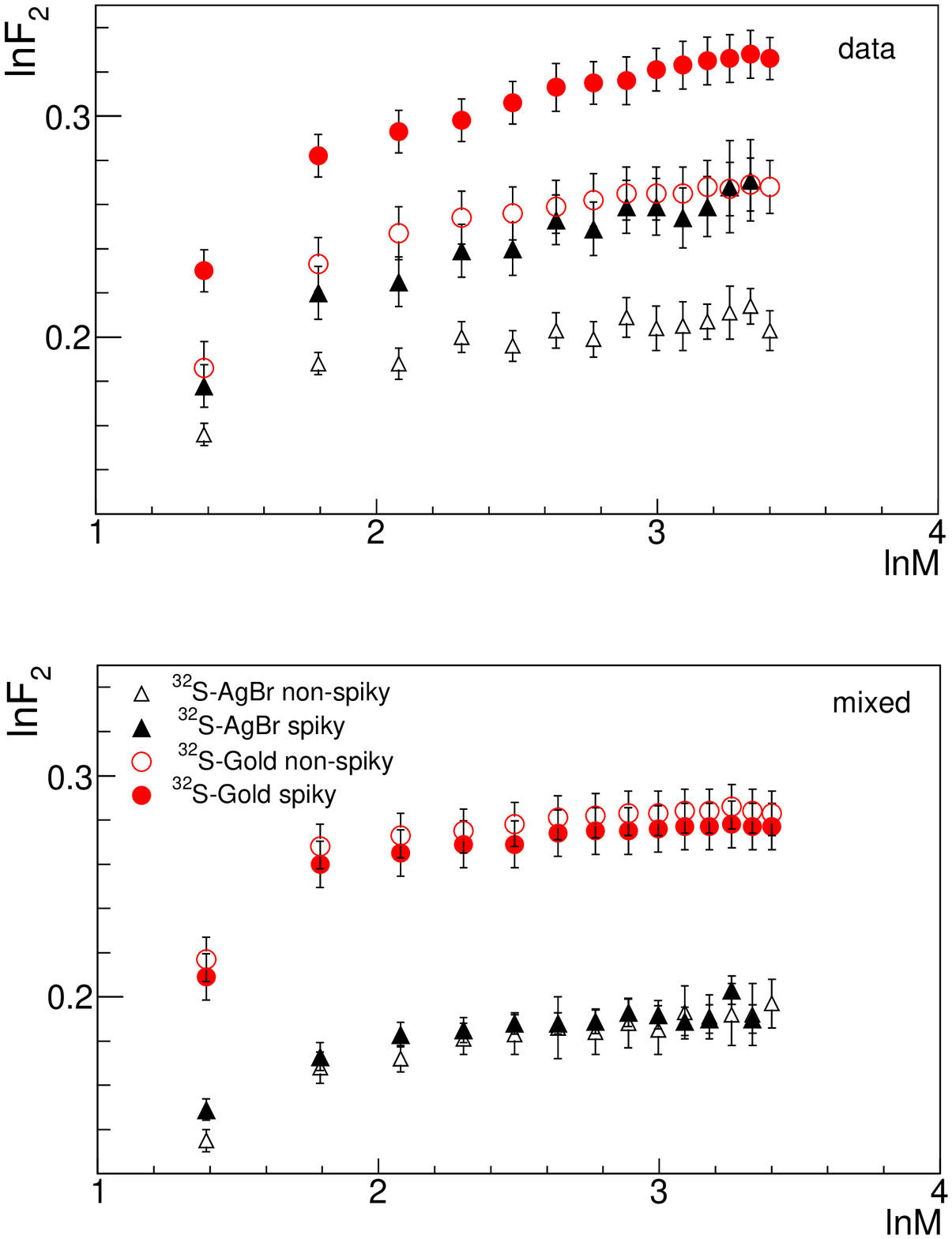,width=10cm}}
\vspace*{8pt}
\caption{Variations of lnF$_2$ with lnM for the experimental and mixed event samples.}
\end{figure}

\newpage
\begin{figure}[th]
\centerline{\psfig{file=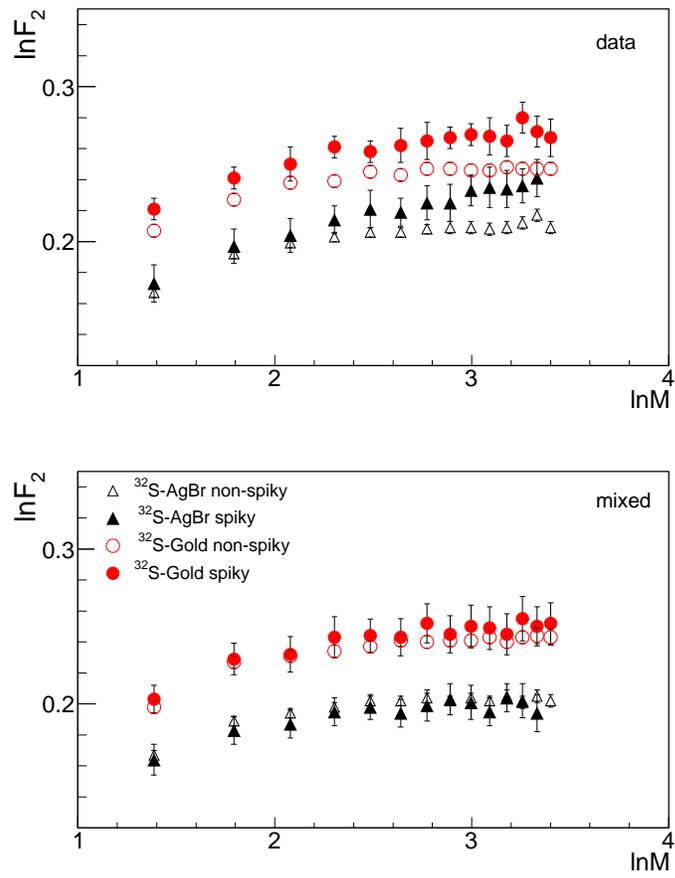,width=10cm}}
\vspace*{8pt}
\caption{Variations of lnF$_2$ with lnM for the \am and corresponding mixed events.}
\end{figure}
\newpage
\begin{figure}[th]
\centerline{\psfig{file=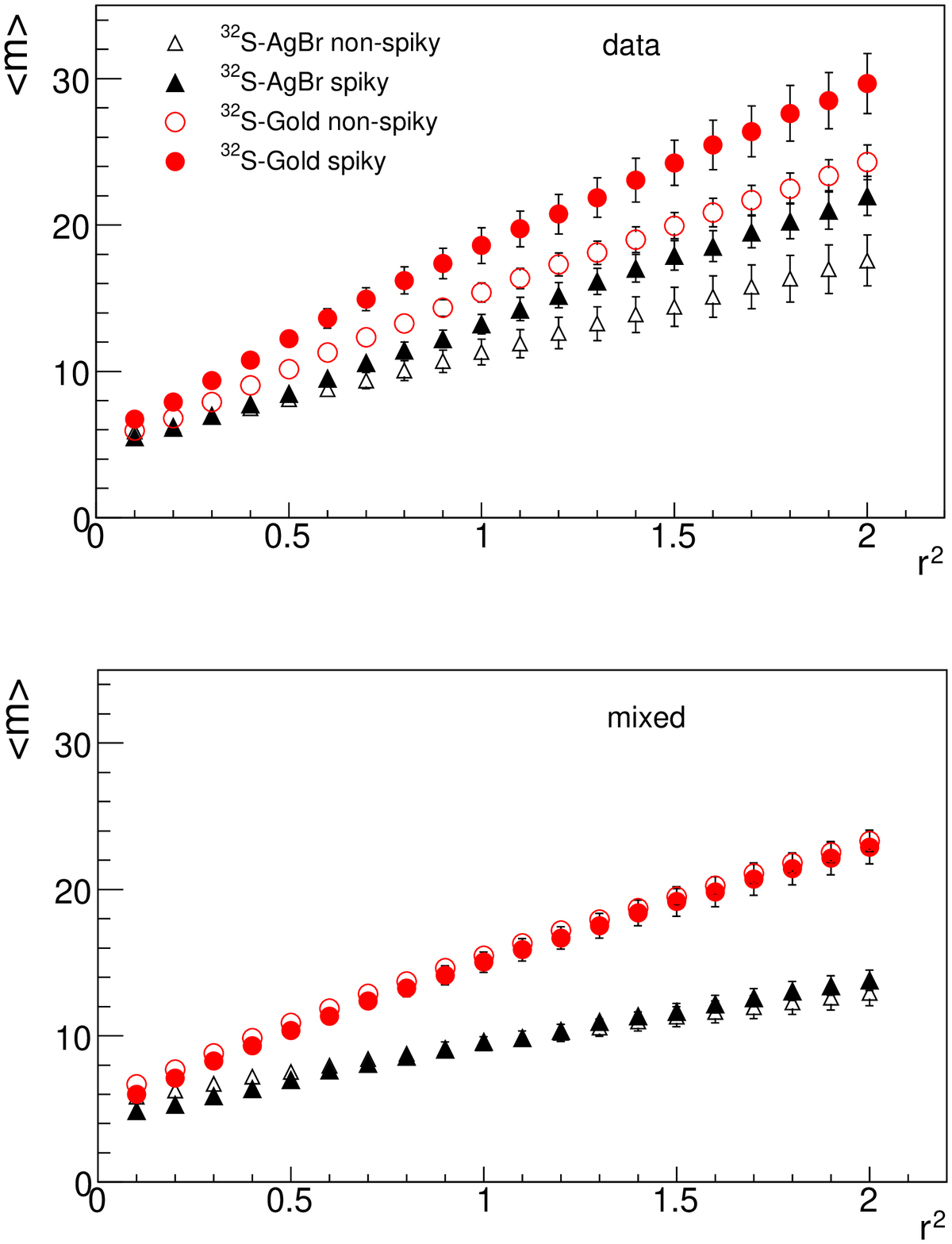,width=10cm}}
\vspace*{8pt}
\caption{Variations of \(<m>\) with \(r^2\) for the experimental and mixed events.}
\end{figure}

\newpage
\begin{figure}[th]
\centerline{\psfig{file=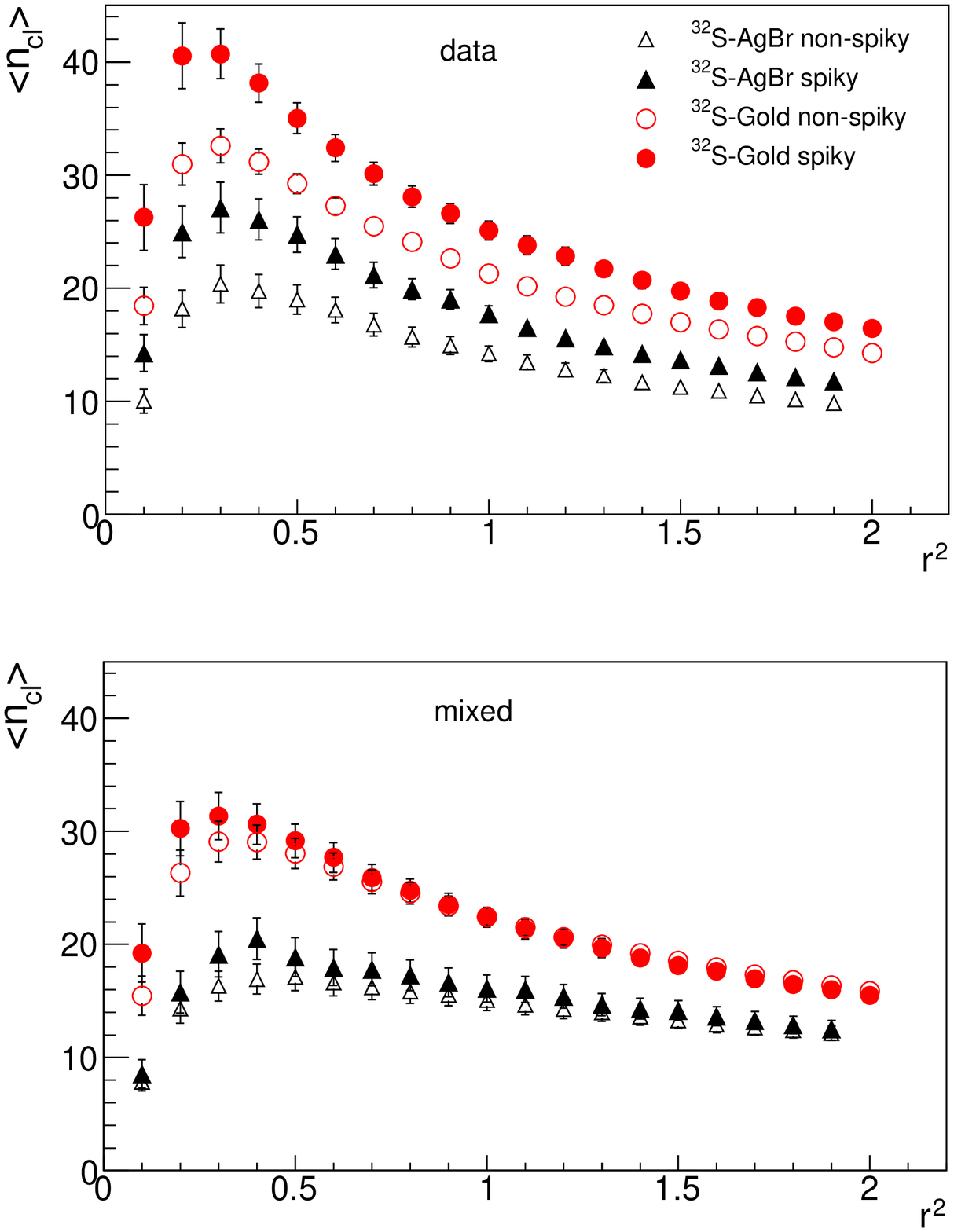,width=10cm}}
\vspace*{8pt}
\caption{\(<n_{cl}>\) vs \(r^2\) plots for the experimental and mixed events. }
\end{figure}

\newpage
\begin{figure}[th]
\centerline{\psfig{file=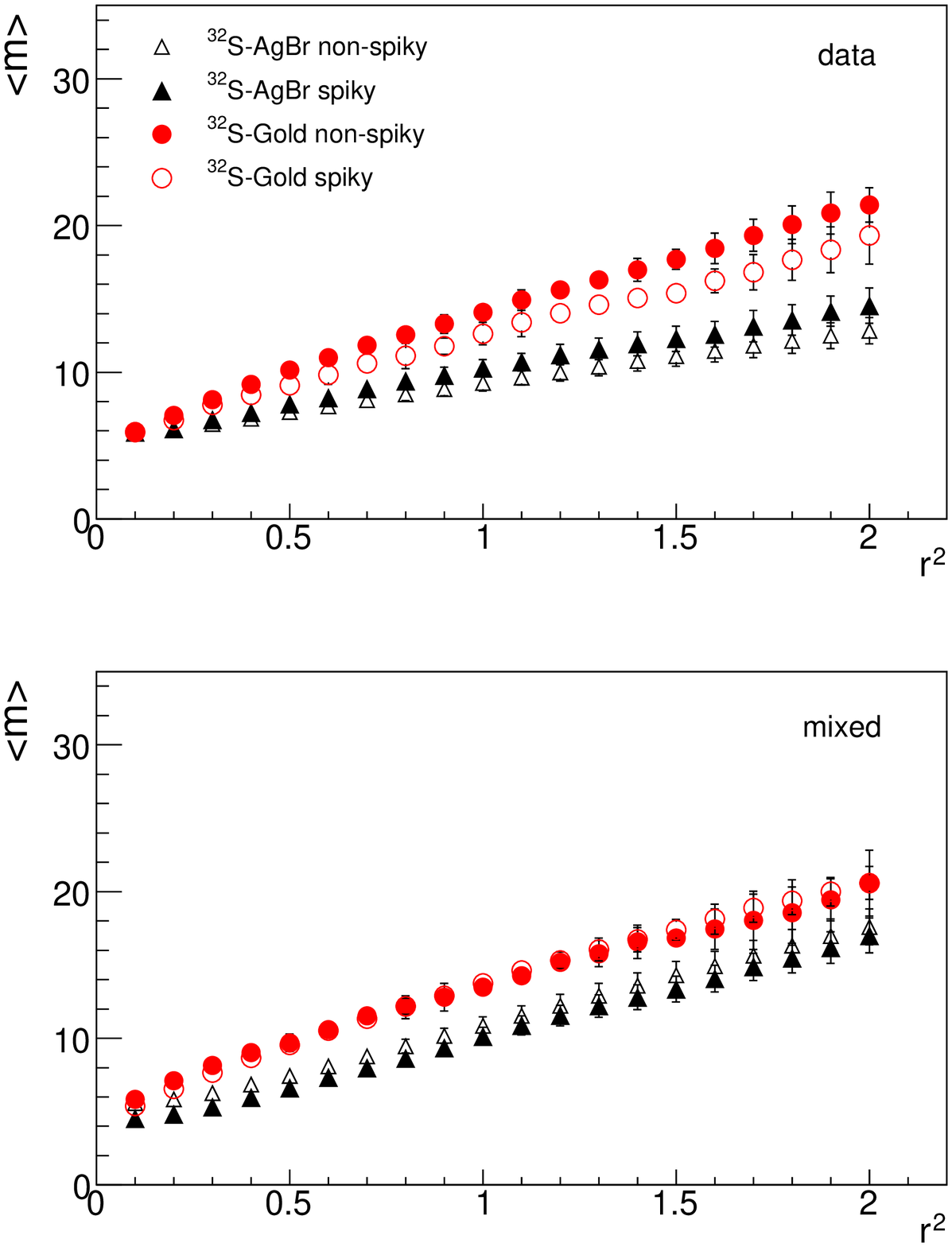,width=10cm}}
\vspace*{8pt}
\caption{Variation of \(<m>\) with \(r^2\) for the \am and mixed event samples.}
\end{figure}
\newpage
\begin{figure}[th]
\centerline{\psfig{file=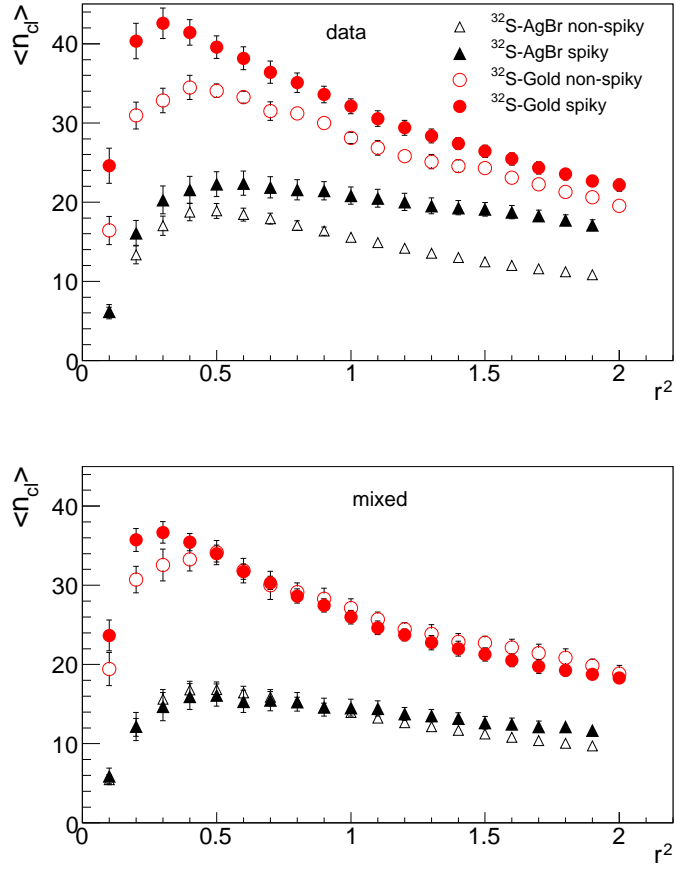,width=10cm}}
\vspace*{8pt}
\caption{\(<n_{cl}>\) vs \(r^2\) plots for the \am and corresponding mixed events.}
\end{figure}
\newpage
\noindent Table~1: Values of mean multiplicities of relativistic charged particles for the experimental and \am generated events.
\begin{footnotesize}
\begin{table}
\begin{center}
\begin{tabular}{|l|l|c|c|} \hline
Type of  & \multicolumn{1}{c|}{Type of} &\(<n_s>_{expt.} \)& \(<n_s>_{AMPT} \) \\
interaction & \multicolumn{1}{c|}{events} & & \\\hline
$^{32}$S-AgBr & all events & 199.43 $\pm$ 6.20 & 205.65 $\pm$ 4.48\\[2mm]
 & (spiky)$_{\Delta \eta = 0.2}$ & 309.32 $\pm$ 17.31 & 276.79 $\pm$ 18.75\\[2mm]
 & (non-spiky)$_{\Delta \eta = 0.2}$ & 188.76 $\pm$ 6.35 & 204.21 $\pm$ 4.62\\[2mm]\hline
$^{32}$S-Gold & all events & 363.28 $\pm$ 4.11 & 357.01 $\pm$ 3.51\\[2mm]
 & (spiky)$_{\Delta \eta = 0.2}$ & 410.36 $\pm$ 7.94 & 368.00 $\pm$ 14.34\\[2mm]
 & (non-spiky)$_{\Delta \eta = 0.2}$ & 350.12 $\pm$ 4.57 & 346.36 $\pm$ 3.54\\[2mm]\hline
\end{tabular}
\end{center}
\end{table}
\end{footnotesize} 


\newpage
\noindent Table~2: Probability(in \%) of occurrence and mean values of \(<d_{ik}>\) for spikes \\ with \dik $\ge$ 2.5 for $^{32}$S-Gold and with $\ge$ 2.2 for $^{16}$S-AgBr interactions.
\begin{footnotesize}
\begin{table}
\begin{center}
\begin{tabular}{|c|c|c|c|c|c|} \hline
Type of & & \multicolumn{2}{|c|}{Expt.} & \multicolumn{2}{|c|}{Mixed} \\ \cline{3-6}
interactions & &   \(\Delta\eta = 0.1\) &  \(\Delta\eta = 0.2\)    &   \(\Delta\eta = 0.1\) &  \(\Delta\eta = 0.2\) \\[2mm] \hline
 $^{32}$S-AgBr & \(P(d_{ik})\) &  0.19  &  0.39  &  0.15  &  0.27  \\[1mm]
     &           \(<d_{ik}>\)  &  2.41 $\pm$ 0.18  &  2.44 $\pm$ 0.22  &  2.34 $\pm$ 0.16  &  2.36 $\pm$ 0.17 \\[2mm] \hline
 $^{32}$S-Gold & \(P(d_{ik})\) &  0.38  &  1.33  & 0.02  &  0.10 \\[1mm]
     &           \(<d_{ik}>\)  &  2.79 $\pm$ 0.25  &  2.98 $\pm$ 0.45  &  2.91    &  2.78 $\pm$ 0.23 \\[2mm] \hline
  & & \multicolumn{2}{|c|}{AMPT} & \multicolumn{2}{|c|}{Mixed} \\ \cline{3-6}
 $^{32}$S-AgBr & \(P(d_{ik})\) &  0.11  &  0.20  &  0.06  &  0.08 \\[1mm]
     &           \(<d_{ik}>\)  &  2.37 $\pm$ 0.18  &  2.45 $\pm$ 0.20  &  2.28 $\pm$ 0.11  & 2.36 $\pm$ 0.12   \\[2mm] \hline
 $^{32}$S-Gold & \(P(d_{ik})\) &  0.06  &  0.08  &  0.04  &  0.06 \\[1mm]
     &           \(<d_{ik}>\)  &  2.65 $\pm$ 0.11  &  2.76 $\pm$ 0.22  &  2.61 $\pm$ 0.08  &  2.65 $\pm$ 0.14 \\[2mm] \hline

\end{tabular}
\end{center}
\end{table}
\end{footnotesize} 


\newpage
\noindent Table~3: Values of \(\phi_2\) for various data sets.
\begin{footnotesize}
\begin{center}
\begin{table}
\begin {tabular}{|c|c|c|c|c|} \hline
& \multicolumn{4}{|c|}{values of \(\phi_2\)}   \\ \hline
Data  & \multicolumn{2}{|c|}{$^{32}$S-AgBr} & \multicolumn{2}{|c|}{$^{32}$S-Gold} \\ \cline{2-5}
type  &   Spiky   &   Non-Spiky   &   Spiky   &   Non-Spiky \\[2mm] \hline
 Expt. & 0.031 $\pm$ 0.003 & 0.017 $\pm$ 0.002   & 0.030 $\pm$ 0.001 & 0.021 $\pm$ 0.002 \\[1mm]
 Mixed & 0.012 $\pm$ 0.002 & 0.015 $\pm$ 0.002   & 0.011 $\pm$ 0.001 & 0.011 $\pm$ 0.001 \\[2mm] \hline

 AMPT  & 0.027 $\pm$ 0.001 & 0.012 $\pm$ 0.001   & 0.019 $\pm$ 0.002 & 0.011 $\pm$ 0.002 \\[1mm]
 Mixed & 0.010 $\pm$ 0.001 & 0.010 $\pm$ 0.001   & 0.014 $\pm$ 0.002 & 0.010 $\pm$ 0.001 \\ [2mm] \hline 
\end{tabular}
\end{table}
\end{center}
\end{footnotesize}

\end{document}